\documentstyle[12pt]{article}
\begin{document}
\begin{titlepage}
\normalsize
\begin{center}
{\Large \bf Budker Institute of Nuclear Physics} 
\end{center}
\begin{flushright}
BINP 96-13\\

March 1996
\end{flushright}
\vspace{0.5cm}
\begin{center}
{\bf NEW CONSTRAINTS FROM ELECTRIC DIPOLE MOMENTS} 
\end{center}
\begin{center}
{\bf ON PARAMETERS}
\end{center}
\begin{center}
{\bf OF THE SUPERSYMMETRIC SO(10) MODEL}
\end{center}

\vspace{0.5cm}

\begin{center}
{\bf I.B. Khriplovich}\footnote{e-mail address: khriplovich@inp.nsk.su}
\end{center}
\begin{center}
Budker Institute of Nuclear Physics, 630090 Novosibirsk, Russia
\end{center}

\begin{center}
{\bf and K.N. Zyablyuk}\footnote{e-mail address: zyablyuk@vitep3.itep.ru}
\end{center}

\begin{center}
Institute of Theoretical and Experimental Physics, 117259 Moscow, Russia
\end{center}

\vspace{1.0cm}

\begin{abstract}
We calculate the chromoelectric dipole moment (CEDM) of $d$- and
$s$-quark in the supersymmetric $SO(10)$ model. CEDM is more
efficient than quark electric dipole moment (EDM), in inducing the
neutron EDM. New, strict constraints on parameters of the
supersymmetric $SO(10)$ model follow in this way from the neutron
dipole moment experiments. As strict bounds are derived from the
upper limits on the dipole moment of $^{199}$Hg.
\end{abstract}

\end{titlepage}

{\bf 1.} The predictions of supersymmetric models for the neutron
electric dipole moment can be comparable with the present
experimental upper limit \cite{sm,al}.  Two different sources of
CP-violation are possible here.  First, the soft breaking potential
contains the mass matrix $m^2_{ij}$ of all scalars and the coupling
matrix $A_{ij}$ of trilinear terms. They are supposed to be defined
by the structure of the hidden sector of some underlying supergravity
and can contain imaginary CP-violating phases.

But the hidden sector can be flavour-blind and CP-invariant. It leads
to the universality condition $m^2_{ij}=\delta_{ij}m^2_0$,
$A_{ij}=\delta_{ij}A_0$ at the Planck scale and to real soft breaking
operators, in particular to $\mbox{Im}A_0=0$. In this situation the
second source of CP-violation becomes essential. In the diagrams of
the type 1a,b each quark-squark-gluino vertex contains the matrix of
the rotation in the generation space of quarks with respect to
squarks. These rotations are different for left- and right-handed
particles, and the $CP$-violating part of the relative rotation
induces the quark EDM. This, second case occurs only in those unified
theories where all quarks in the given generation belong to the same
representation of the unification group \cite{dh}. This is the case
for the supersymmetric SO(10) (but not for MSSM and supersymmetric
SU(5).

The neutron EDM, as induced by the quark EDM, in the supersymmetric
SO(10) was considered in Ref. \cite{dh}. The same model is discussed
in the present note. But as distinct from Ref. \cite{dh}, we
concentrate on another $CP$-odd characteristic of a quark, its CEDM,
and effects induced by it.

\bigskip

{\bf 2.} The quark CEDM is defined as the factor $d^c$ in the effective
operator
\begin{equation}\label{eo}
L_{eff}=\,\frac{1}{2}\,d^c\,\bar q \gamma_5\sigma_{\mu\nu}\frac{\lambda^a}{2}
q\,G^a_{\mu\nu}.
\end{equation}
It is generated by diagrams 1a,b (obviously, only diagrams of the
type 1a contribute to the quark EDM).

As usual, we choose the Yukawa coupling matrix $\lambda^U$ of
$U$-quarks ($U=u,c,t$) real and diagonal. Then one can take into
account in the renormalization group equations, written in Ref.
\cite{bhs}, the top Yukawa coupling only. In this situation the
$U$-quarks do not rotate in the generation space with respect to
their superpartners. Therefore, in the case considered, of
CP-invariant soft breaking operators, both EDMs and CEDMs of
$U$-quarks are negligible.

The calculations of the $d$-quark EDM $d$ and CEDM $d^c$ are quite
similar. If all squarks were degenerate as at the Planck scale,
their contributions would cancel. However, due to large top Yukawa
coupling, the third generation becomes considerably lighter already
at the GUT scale \cite{dh,bh}. So, we take into account only
contribution of the $b$-squark as the largest one.  The result of
both EDM and CEDM calculations can be conveniently presented as
\begin{equation}\label{de}
d\,=\,e\,\frac{\alpha_s}{54\pi}\,{v_d \over m^3_B}\,
f\left({{\tilde m} \over m_B}\right) \mbox{Im} \left[\,(V_L)_{31}\,
(V^*_R)_{31}\,(A^D\lambda^D + \mu\lambda^D \tan\beta)_{33}\,\right];
\end{equation}
\begin{equation}
\label{dce}
d^c\,=\,g\,{5\alpha_s \over 72\pi}\,{v_d \over m^3_B}\,
f^c\left({{\tilde m} \over m_B}\right) \mbox{Im} \left[\,(V_L)_{31}\,
(V^*_R)_{31}\,(A^D\lambda^D + \mu\lambda^D \tan\beta)_{33}\,\right].
\end{equation}
Here $\mu$ is the constant of the $\mu H_1 H_2$ superpotential,
$\tan\beta=v_u/v_d$ is the ratio of vacuum expectation values of
$H_2$ and $H_1$; $A^D$ is the $3 \times 3$-matrix in the trilinear
soft breaking potential. We neglect the splitting between the masses
$m_B$ of left- and right-handed $b$-squarks.  The dependence of the
dipole moments on the gluino mass $\tilde m$ is determined by the
functions $f$ and $f^c$:
\begin{equation}
f(x)\,=\,6\,x\,{1+5\,x^2 \over (1-x^2)^3} \,+\,
24\,x^3\,{2+x^2 \over (1-x^2)^4}\,\ln x\,; 
\end{equation}
\begin{equation}
f^c(x)\,=\,-\,12\,x\,{11+\,x^2 \over 5\,(1-x^2)^3} \,-\,
12\,x\,{9\,+\,16\,x^2\,-\,x^4 \over 5\,(1-x^2)^4}\,\ln x\,.
\end{equation}
Here $x\,=\,\tilde m/m_B$, and both functions, $f(x)$ and
$f^c(x)$ are normalized in such a way that $f(1)\,=\,f^c(1)\,=\,1$.

$V_L$ and $V_R$ are the matrices of the unitary rotations of
left- and right-handed quarks with respect to their superpartners.
They are related to the Yukawa coupling matrix $\lambda^D$ as
follows:
\begin{equation}
\lambda^D = \frac{1}{v_d}\,V^*_L \,{\bar M}^D\, V^T_R \;,
\end{equation}
where ${\bar M}^D$ is the real diagonal mass matrix of
$D$-quarks ($d,\,s,\,b$).

$V_L$ is nothing else but the Kobayashi-Maskawa matrix $V$.
Meanwhile, in the MSSM and supersymmetric $SU(5)$ model $V_R$ is the
unit matrix and the dipole moments vanish \cite{dh}.  However, in the
supersymmetric $SO(10)$ model 
$$V^*_R=V P^2,$$ 
where $P$ is a diagonal
phase matrix with two physical phases \cite{bhs1}. In this model the
$d$-quark dipole moments, generally speaking, do not vanish. They can
be written as
\begin{equation}\label{edm}
d\,=\,e\,{\alpha_s \over 54\pi}\,\vert V_{31} \vert^2\,A'_b\,\sin\phi\,
{m_b \over m^2_B}\,f\left({{\tilde m} \over m_B}\right);
\end{equation}
\begin{equation}\label{cedm}
d^c=\,g_s\,{5\alpha_s \over 72\pi}\,\vert V_{31} \vert^2\,A'_b\,\sin\phi\,
{m_b \over m^2_B}\,f^c\left({{\tilde m} \over m_B}\right).
\end{equation}
In these expressions $m_b$ is the b-quark mass,
$$A'_b={A^D_{33}+\mu\tan\beta \over m_B}.$$  
The phase $\phi$ is the
sum over all phases present in (\ref{de}) and (\ref{dce}):
$$\phi=2\,(\phi_{31}-\phi_{33}+{\tilde \phi}_1-{\tilde \phi}_3),$$
where $\phi_{ij}\,=\,\arg V_{ij}$, $P_{ii}=e^{i{\tilde
\phi}_i}$. At $\tilde m\,=\, m_B$ formula (\ref{edm}) for the
$d$-quark EDM coincides with the result obtained in Ref. \cite{dh}.

The constraints put on the parameters of the supersymmetric $SO(10)$
model by the neutron dipole moment, as induced by the $d$-quark EDM
(\ref{edm}), were considered in Ref. \cite{dh}. We will discuss here
the constraints following from our CEDM result (\ref{cedm}).

But before of that we wish to mention the following circumstance. Let
us consider, instead of the vertex part (Fig. 1), the corresponding
mass operator (Fig. 2). This contribution to the $CP$-odd
$\gamma_5$-mass of a quark, or to the induced $\theta$-term, is 
enormous. It exceeds
by 6 -- 7 orders of magnitude the upper limit on $\theta$ \cite{bal,cdv}
following from the neutron EDM experiment. This situation is quite
common to models of $CP$-violation. As common is the argument,
according to which there should be some mechanism, for instance the
Peccei-Quinn one, which makes the $\theta$-term harmless, which
allows to transform it away. We will also adhere to this conservative
point of view. 

\bigskip

{\bf 3.} Coming back to the quark CEDM, to investigate its
contribution to the observable effects, we have to bring the
expression (\ref{cedm}) down from the scale of $M\,\sim\,300$ GeV. In
particular, to substitute for $m_b$ its "physical" value $4.5$ GeV,
we have to introduce the renormalization group (RG) factor
$$\left[\frac{\alpha_s(M)}{\alpha_s(m_b)}\right]^{12/23}.$$ 
Now, the QCD sum rule technique, used below to estimate the CEDM
contribution to observable effects, refers to the hadronic scale of
$m\,\sim\,1$ GeV and is applied directly to the operators of the type
$$g_s\,\bar q\,\gamma_5\sigma_{\mu\nu}\frac{\lambda^a}{2}\,q\,G^a_{\mu\nu},$$
which include $g_s$ explicitly. This brings one more RG factor
\cite{svz} $$\left[\frac{\alpha_s(M)}{\alpha_s(m)}\right]^{2/23}.$$

On the other hand, as distinct from some other investigations, we see
no special reasons to bring the explicit $\alpha_s$ factor, entering
the expression (\ref{cedm}), down from the high-momenta scale $M$,
where it is defined at least as well as at $m\,\sim\,1$ GeV.

The overall RG factor, introduced in this way into formula
(\ref{cedm}), is
\begin{equation}
\left[\frac{\alpha_s(M)}{\alpha_s(m_b)}\right]^{12/23}
\left[\frac{\alpha_s(M)}{\alpha_s(m)}\right]^{2/23}\,=\,0.57.
\end{equation}
The values of the coupling constants, accepted here, are:
$$\alpha_s(M)\,=\,0.11;\;\;\;\; \alpha_s(m_b)\,=\,0.26;\;\;\;\; 
\alpha_s(m)\,=\,0.43.$$

If, following \cite{dh}, we assume for the estimates $\tilde
m\,=\,m_B$, then at the same, as in Ref. \cite{dh}, representative values of
other parameters, the $d$-quark CEDM can be evaluated as follows:
\begin{equation}\label{prc}
d^c=\,26 \cdot 10^{-26}\,\mbox{cm}\,\left(\frac{|V_{td}|^2}{10^{-4}}\right)\,
\left(\frac{A^{\prime}}{1}\right)\,\left(\frac{\sin \phi}{0.5}\right)
\left(\frac{250\mbox{GeV}}{m_B}\right)^2.
\end{equation}

A serious problem is to find the CEDM contribution to the
neutron dipole moment. The simplest way \cite{kk} to estimate this
contribution is to assume, just  by dimensional reasons, that
$d(n)/e$ is roughly equal to $d^c(q)$ (obviously, the electric charge
$e$ should be singled out of $d(n)$, being a parameter unrelated to
the nucleon structure).

In a more elaborate approach \cite{kk}, the CEDM contribution to the 
neutron EDM is estimated in the chiral limit via diagram 3 (see Ref.
\cite{cdv}). The contribution of operator (\ref{eo}) to the $CP$-odd 
$\pi NN$ constant $\bar{g}_{\pi NN}$ is transformed by the PCAC
technique:
\begin{equation}\label{pi}
<\,\pi^- p\,|\,g_s\,\bar q \gamma_5\sigma_{\mu\nu}\frac{\lambda^a}{2}\,
q\,G^a_{\mu\nu}|\,n\,> = \, \frac{i}{f_{\pi}}\,
 <\,p\,|\,g_s\,\bar u \sigma_{\mu\nu}\frac{\lambda^a}{2}\,d\,
 G^a_{\mu\nu}\,|\,n\,>. 
\end{equation}
QCD sum rule estimate gives for the last matrix element value close 
to\newline  
$-1.5$ GeV$^2$. Let us introduce now the ratio of the neutron dipole 
moment, as induced by a CEDM, to $d^c$ itself:
\begin{equation}\label{ro}
\rho\,=\,\frac{d(n)/e}{d^c}.
\end{equation}
Its value obtained in this, more elaborate approach, 
$$\rho = 0.7,$$ 
is quite close indeed to unity. In our opinion, this good agreement with
the above simple-minded result enhances the reliability of both
estimates.

In this way at $\rho = 0.7$ we obtain the following prediction for the
neutron EDM: 
\begin{equation}\label{prd}
d(n)/e\,=\,18 \cdot 10^{-26}\,\mbox{cm}\,
\left(\frac{|V_{td}|^2}{10^{-4}}\right)\,
\left(\frac{A^{\prime}}{1}\right)\,\left(\frac{\sin \phi}{0.5}\right)
\left(\frac{250\mbox{GeV}}{m_B}\right)^2.
\end{equation}
It should be compared with the the experimental upper limit
\cite{sm,al}
\begin{equation}\label{exp}
d(n)/e\,<\,7 \cdot 10^{-26}\,\mbox{cm}.
\end{equation}
The prediction (\ref{prd}) for the neutron dipole moment, as induced
by the quark CEDM, is 4 times larger than the contribution to $d(n)$
from the quark EDM \cite{dh}. Correspondingly, it constrains stronger
the parameters of the supersymmetric $SO(10)$ model. 

\bigskip

{\bf 4.} Essentially larger contribution to the neutron EDM is
induced by the CEDM $d^c(s)$ of the $s$-quark. The expression for
$d^c(s)$ differs from (\ref{prc}) in two respects. First, the
concrete expression for the phase $\phi$ changes. But what is 
more essential, the mixing between the second and third generations
is essentially larger than the mixing between the first and third
ones: $$|V_{ts}|^2\,\simeq\,17\cdot 10^{-4}.$$ (Let us mention that
in other models the advantage of the $s$-quark contribution is the
large mass ratio $m_s/m_d$ \cite{hmp}.)

On the other hand, for the $s$-quark, the ratio 
\begin{equation}\label{ros}
\rho_s\,=\,\frac{d(n)/e}{d^c(s)}.
\end{equation}
should be much smaller than unity. Indeed, according to the QCD sum
rule calculations of \cite{kkz}, it is about 0.1. One should mention
that other estimates \cite{zk,hmp1} predict for the ratio (\ref{ros})
a value an order of magnitude smaller. 

Then, how reliable is the estimate
$\rho_s\,=\,0.1$ ? There are strong indications now that the
admixture of the $\bar s s$ pairs in nucleons is quite considerable.
In particular, it refers to the spin content of a nucleon. And though
these indications refer to operators different from 
$\bar s \gamma_5\sigma_{\mu\nu}\,(\lambda^a/2)\,s\,G^a_{\mu\nu}$,
they give serious reasons to believe that the estimate
\begin{equation}
\rho_s\,=\,0.1
\end{equation}
is just a conservative one.

The central point of the contribution of the $s$-quark CEDM
to the neutron dipole moment, resulting at $\rho_s\,=\,0.1$,
\begin{equation}\label{}
d(n)/e\,=\,43 \cdot 10^{-26}\,\mbox{cm}\,
\left(\frac{A^{\prime}}{1}\right)\,\left(\frac{\sin \phi_s}{0.5}\right)
\left(\frac{250\mbox{GeV}}{m_B}\right)^2.
\end{equation} 
is 6 times larger than the experimental upper limit (\ref{exp}).

\bigskip

{\bf 5.} Let us compare at last the predictions of the supersymmetric $SO(10)$
model with the result of the atomic experiment \cite{lam}. The
measurements of atomic EDM of the mercury isotope $^{199}$Hg have
resulted in
\begin{equation}\label{hg}
d(^{199}\mbox{Hg})/e\,<\,9\cdot 10^{-28}\mbox{cm}.
\end{equation}
According to calculations of Ref. \cite{kky}, it corresponds to the
upper limit on the $d$-quark CEDM 
\begin{equation}\label{}
d^c\,<\,2.4\cdot 10^{-26}\mbox{cm}
\end{equation}
The central point of the prediction (\ref{prc}) exceeds this upper
limit by an order of magnitude.

The analysis carried out in the present paper demonstrates that very
special assumptions concerning the parameters of the supersymmetric
$SO(10)$ model (such as large mass $m_B$ of the $b$-squark, small
$CP$-violating angle $\phi$, etc) are necessary to reconcile the
predictions of this model with the experimental upper limits on the
electric dipole moments of neutron and $^{199}$Hg.

\bigskip
\bigskip
\bigskip
\bigskip

We are extremely grateful to M.E. Pospelov and M.V. Terentjev for
their interest to the work and helpful discussions. We thank also 
J. Ellis for the discussion of results. The investigation is
supported by the Russian Foundation for Basic Research
through grants No.95-02-04436-a and 95-02-05822-a.
One of us (I.B. Kh.) acknowledges support by the National Science
Foundation through a grant to the Institute for Theoretical Atomic
and Molecular Physics at Harvard University and Smithsonian
Astrophysical Observatory. 

\newpage
Figure captions:

\bigskip

Fig. 1. Gluino contributions to quark dipole moments.
$B_{L(R)}$ denotes the left(right)-handed $b$-squark.

\bigskip

Fig. 2. Gluino contribution to the $\theta$-term.

\bigskip

Fig. 3. Chiral contribution to the neutron EDM.
$\pi NN$ vertices {\bf 1} and
i$\gamma_5$ refer to the $CP$-odd and usual strong interactions, 
respectively. 


\newpage

\begin{thebibliography}{99}
\bibitem{sm} K.F. Smith et al, Phys.Lett. B 234 (1990) 191
\bibitem{al} I.S. Altarev et al, Phys.Lett. B 276 (1992) 242
\bibitem{dh} S.Dimopoulos, L.J.Hall, Phys.Lett. B 344 (1995) 185
\bibitem{bhs} R.Barbieri, L.Hall, A.Strumia, Nucl.Phys B 445 (1995) 219
\bibitem{bh} R.Barbieri, L.J.Hall, Phys.Lett. B 338 (1994) 212
\bibitem{bhs1} R.Barbieri, L.Hall, A.Strumia, Nucl.Phys B 449 (1995) 437
\bibitem{bal} V. Baluni, Phys.Rev. D 19 (1979) 2227
\bibitem{cdv} R.J.Crewter, P.Di Veccia, G.Veneziano, E.Witten,
Phys.Lett. B 88 (1979) 123; B 91 (1980) 487 (E)
\bibitem{svz} M.A. Shifman, A.I. Vainshtein, V.I. Zakharov,
Phys.Rev. D 77 (1978) 2583
\bibitem{kk} V.M. Khatsymovsky, I.B. Khriplovich, Phys.Lett. B 296
(1992) 219
\bibitem{hmp} X.-G. He, B.H.J. McKellar, S. Pakvasa, Phys.Lett. B
254 (1991) 231
\bibitem{kkz} V.M. Khatsymovsky, I.B. Khriplovich, A.R. Zhitnitsky, 
Z.Phys. C 36 (1987) 455
\bibitem{zk} A.R. Zhitnitsky, I.B. Khriplovich, Yad.Fiz. 34 (1981)
167\newline [Sov.J.Nucl.Phys. 34 (1982)]
\bibitem{hmp1} X.-G. He, B.H.J. McKellar, S. Pakvasa, Int.J.Mod.Phys. A 
4 (1989) 5011
\bibitem{lam} J.P. Jacobs, W.M. Klipstein, S.K. Lamoreaux, B.R. Heckel, E.N.
Fortson, Phys.Rev. A 52 (1995) 3521
\bibitem{kky} V.M. Khatsymovsky, I.B. Khriplovich, A.S. Yelkhovsky, 
Ann.Phys. 186 (1988) 1

\end{thebibliography}
\end{document}